\documentclass[%
 reprint,
showpacs,
bibnotes,
 amsmath,amssymb,
 aps,
floatfix,
altaffilletter
]{revtex4-1}

\usepackage{graphicx}
\usepackage{dcolumn}
\usepackage{bm}

\begin{document}

\preprint{APS/123-QED}

\title{Transition to chaos in random networks with cell-type-specific connectivity}
\author{Johnatan Aljadeff$^{1,2,\text{a}}$}
\author{Merav Stern$^{3,4}$}
\thanks{Equal contribution}
\author{Tatyana O. Sharpee$^{1,2,}$}
\email{sharpee@salk.edu}
\affiliation{$^1$Computational Neurobiology Laboratory, The Salk Institute for Biological Studies, La Jolla, California, USA}
\affiliation{$^2$Center for Theoretical Biological Physics and Department of Physics, University of California, San Diego, USA}
\affiliation{$^3$Department of Neuroscience, Columbia University, New York, New York, USA}
\affiliation{$^4$The Edmond and Lily Safra Center for Brain Sciences, Hebrew University, Jerusalem, Israel}

\date{\today}

\begin{abstract}
In neural circuits, statistical connectivity rules strongly depend on neuronal type. Here we study dynamics of neural networks with cell-type specific connectivity by extending the dynamic mean field method, and find that these networks exhibit a phase transition between silent and chaotic activity. By analyzing the locus of this transition, we derive a new result in random matrix theory: the spectral radius of a random connectivity matrix with block-structured variances. We apply our results to show how a small group of hyper-excitable neurons within the network can significantly increase the network's computational capacity.
\end{abstract}


\maketitle

Conventional firing-rate models used to describe irregular activity in neural networks assume that connections between neurons follow a single connectivity rule. In contrast to this class of models, recent experiments highlight the diversity among neuron types, each having a different degree of excitability and different form of connectivity \cite{Schubert2003,Yoshimura2005,Suzuki2012,Franks2011,Levy2012}. As a step towards bridging this gap between theory and experiment, we extend the conventional firing-rate models and the mean-field methods used to analyze them to the case of multiple cell-types and allow for cell-type-dependent connectivity.

We show that multiple cell-type networks, like networks with a single cell-type \cite{Sompolinsky1988}, exhibit a phase transition between silent and chaotic activity. Previous studies suggest that these networks have optimal computational capacity close to the critical point at which this tranistion occurs  \cite{Bertschinger2004,Toyoizumi2011}. In networks with cell-type specific connectivity the transition depends on the network's connectivity structure, and is related to the spectral properties of the random connectivity matrix serving as the model network's connectivity matrix. Specifically, by finding the location of the critical point for the multiple cell-type network we derive a new result in random matrix theory: the support of the spectral density of asymmetric matrices with block-structured variances.

We also show that the dynamical mean field equations provide predictions for the autocorrelation modes that can be concurrently sustained by a multiple cell-type network. Finally, we apply our results to a network that includes a small group of hyper-excitable neurons, and explain how this small group can significantly increase the network's computational capacity by bringing it into the chaotic regime.

\paragraph*{The critical point.}
The starting point for our analysis of recurrent activity in neural networks is a firing-rate model where the activation $x_i(t)$ of the $i$th neuron
determines its firing-rate $\phi_i(t)$ through a nonlinear function $\phi_i(t) = \tanh(x_i)$. The activation of the $i$th neuron depends on the firing-rate of all $N$ neurons in the network:
\begin{equation}
    \dot{x}_i(t) = -x_i(t) + \sum_{j = 1}^N J_{ij}\phi_j(t),
    \label{eq:dyn0}
\end{equation}
where $J_{ij}$ describes the connection weight from neuron $j$ to $i$. Previous work \cite{Sompolinsky1988} considered a recurrent random network where all connections are drawn from the same distribution, grouping all neurons into a single cell-type. In that work the distribution of matrix elements was chosen to be Gaussian with mean zero and variance $g^2/N$, where the parameter $g$ defines the average synaptic gain in the network. According to Girko's circular law, the spectral density of the random matrix $\mathbf{J}$ in this case is uniform on a disk with radius $g$ \cite{Girko1984,Bai1997}. When the real part of some of the eigenvalues of $\mathbf{J}$ exceeds $1$, the quiescent  state $x_i(t)=0$ becomes unstable and the network becomes chaotic \cite{Sompolinsky1988}. Thus, for networks with cell-type independent connectivity the transition to chaotic dynamics occurs when $g=1$.

\begin{figure*}
\begin{centering}
\includegraphics[width=0.95\textwidth]{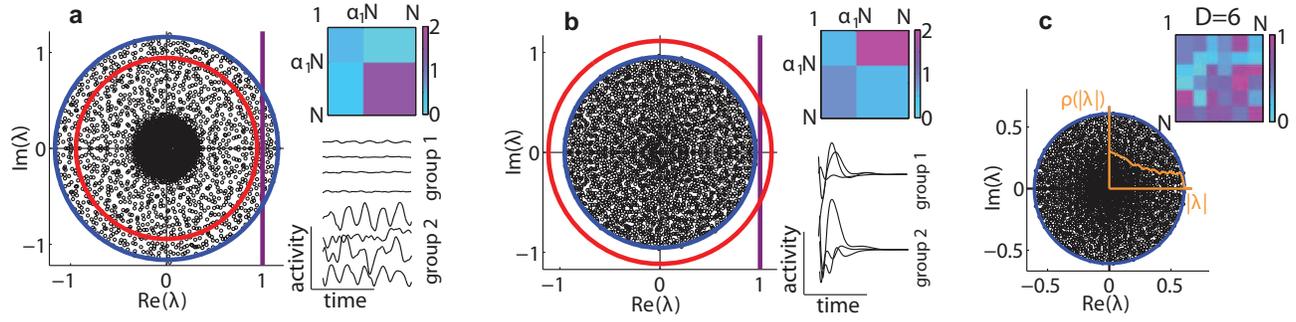}
    \caption{Spectra and dynamics of networks with cell-type dependent connectivity ($N = 2500$). The support of the spectrum of the connectivity matrix $\mathbf{J}$ is accurately described by $\sqrt{\Lambda_1}$ (radius of blue circle) for different networks. Top insets - the synaptic gain matrix $\mathbf{G}$ summarizes the connectivity structure. Bottom insets - activity of representative neurons from each type. The line $\Re\{\lambda\} = 1$ (purple) marks the transition from quiescent to chaotic activity.  (a) An example of chaotic network with two cell types. The average synaptic gain $\bar{g}$ (radius of red circle) incorrectly predicts this network to be quiescent. (b) An example silent network. The average synaptic gain $\bar{g}$ incorrectly predicts this network to be chaotic. (c) An example network with six cell-types. In all examples the radial part of the eigenvalue distribution $\rho(|\lambda|)$ (orange line) is not uniform.}
    \label{fig:example}
\end{centering}
\end{figure*}

We extend these results to networks with $D$ cell-types, where each cell-type (or group of neurons) has a fraction $\alpha_d$ of neurons in it. The mean connection weight is $\left\langle J_{ij}\right\rangle = 0$. The variances $N\langle J^2_{ij}\rangle = g^2_{c_i d_j}$ depend on the cell-type of the input ($c$) and output ($d$) neurons; where $c_i$ denotes the group neuron $i$ belongs to. In what follows, indices $i,j = 1,\dots,N$ and $c,d = 1,\dots,D$ correspond to single neurons and neuron groups, respectively. Averages over realizations of $\mathbf{J}$ are denoted by $\left\langle \cdot \right\rangle$. It is convenient to represent the connectivity structure using a synaptic gain matrix $\mathbf{G}$. Its elements $G_{ij} = g_{c_i d_j}$ are arranged in $D^2$ blocks of sizes $N \alpha_c \times N \alpha_d$ (Fig. 1a-c, top insets). The mean synaptic gain, $\bar{g}$, is given by $N^{-1}(\sum_{i,j=1}^N G^2_{ij})^{\frac{1}{2}} = (\sum_{c,d=1}^D\alpha_c\alpha_d g_{cd}^2)^{\frac{1}{2}}$.

Defining $J^0_{ij} \sim \mathcal{N}\left(0,N^{-1}\right)$ and $n_d = N\sum_{c = 1}^d\alpha_c$ allows us to rewrite Eq.~(\ref{eq:dyn0}) in a form that emphasizes the separate contributions from each group to a neuron:
\begin{equation}
\dot{x}_i = -x_i + \sum_{d = 1}^D g_{c_i d}\sum_{j = n_{d - 1} + 1}^{n_d}J^0_{ij}\phi_j\left(t\right).
\label{eq:dyn}
\end{equation}

We use the dynamic mean field approach \cite{Amari1972,Sompolinsky1982,Sompolinsky1988} to study the network behavior in the $N\rightarrow\infty$ limit. Averaging Eq.~(\ref{eq:dyn}) over the ensemble from which $\mathbf{J}$ is drawn implies that only neurons that belong to the same group are statistically identical. Therefore, to represent the network behavior it is enough to look at the activities $\xi_d(t)$ of $D$ representative neurons and their inputs $\eta_d\left(t\right)$.

The stochastic mean field variables $\xi$ and $\eta$ will approximate the activities and inputs in the full $N$ dimensional network provided that they satisfy the dynamic equation
\begin{equation}
\dot{\xi}_d\left(t\right) = -\xi_d\left(t\right)+\eta_d\left(t\right),
\label{eq:mfield}
\end{equation}
and provided that  $\eta_d\left(t\right)$ is drawn from a Gaussian distribution with moments satisfying the following conditions. First, the mean $\langle\eta_d(t)\rangle = 0$ for all $d$. Second, the correlations of $\eta$ should match the input correlations in the full network, averaged separately over each group. Using Eq. (\ref{eq:mfield}) and the property $N\left\langle J^0_{ij}J^0_{kl} \right\rangle = \delta_{ik} \delta_{jl}$ we get the self-consistency conditions:
\begin{widetext}
\begin{equation}
\left\langle\eta_c\left(t\right)\eta_d\left(t+\tau\right)\right\rangle = \sum_{a, b = 1}^D\sum^{n_a}_{j = n_{a - 1} + 1} \sum^{n_b}_{l = n_{b - 1} + 1} g_{c_i a} g_{d_j b} \left\langle J^0_{ij}J^0_{kl} \right\rangle \left\langle\phi\left[x_j(t)\right] \phi\left[x_l(t+\tau)\right] \right\rangle = \delta_{cd}\sum_{b = 1}^D\alpha_b g^2_{cb}C_b(\tau),
\label{eq:mf_constraints}
\end{equation}
\end{widetext}
where $\mathbf{C}\left(\tau\right)$ is the average firing rate correlation vector. Its components (using the variables of the full network and averaging over time) are $C_d(\tau)=\sum^{n_d}_{i=n_{d-1}+1} \left\langle\phi[x_i(t)]\phi[x_i(t+\tau)]\right\rangle$, translating to $C_d (\tau) =
\left\langle\phi[\xi_d(t)]\phi[\xi_d(t+\tau)]\right\rangle$ using the mean field variables. Importantly, the covariance matrix $\bm{\mathcal{H}}(\tau)$ with elements $\mathcal{H}_{cd}\left(\tau\right)=\left\langle\eta_c\left(t\right)\eta_d\left(t+\tau\right)\right\rangle$ is diagonal, justifying the definition of the vector $\mathbf{H}=\text{diag}\left(\bm{\mathcal{H}}\right)$. With this in hand we rewrite Eq.~(\ref{eq:mf_constraints}) in matrix form as
\begin{equation}
    \mathbf{H}\left(\tau\right) = \mathbf{MC}\left(\tau\right),
    \label{eq:matrix_form}
\end{equation}
where $\mathbf{M}$ is a constant matrix reflecting the network connectivity structure: $M_{cd} = \alpha_d g^2_{cd}$.

A trivial solution to this equation is $\mathbf{H}(\tau) = \mathbf{C}(\tau) = 0$ which corresponds to the silent network state: $x_i(t) = 0$. Recall that in the network with a single cell-type, the matrix $\mathbf{M} = g^2$ is a scalar and Eq.~(\ref{eq:matrix_form}) reduces to $H(\tau) = g^2 C(\tau)$. In this case the silent solution is stable only when $g<1$. For $g>1$ the autocorrelations of $\eta$ are non-zero which leads to chaotic dynamics in the $N$ dimensional system \cite{Sompolinsky1988}. 

In the general case ($D\ge 1$), Eq.~(\ref{eq:matrix_form}) can be projected on the eigenvectors of $\mathbf{M}$ leading to $D$ consistency conditions, each equivalent to the single group case. Each projection has an effective scalar given by the eigenvalue in place of $g^2$ in the $D=1$ case. Hence, the trivial solution will be stable if all eigenvalues of $\mathbf{M}$ have real part $<1$. This is guaranteed if $\Lambda_1$, the largest eigenvalue  of $\mathbf{M}$, is $<1$ \footnote{Note that $\mathbf{M}$ has strictly positive elements, so by the Perron-Frobenious theorem its largest eigenvalue (in absolute value) is real and positive and the corresponding eigenvector has strictly positive components.}. If $\Lambda_1>1$ the projection of Eq.~(\ref{eq:matrix_form}) on the leading eigenvector of $\mathbf{M}$ gives a scalar self-consistency equation analogous to the $D=1$ case for which the trivial solution is unstable. As we know from the analysis of the single cell-type network, this leads to chaotic dynamics 
in the full network. Therefore $\Lambda_1=1$ is the critical point of the multiple cell-type network.

Another approach to show explicitly that $\Lambda_1=1$ at the critical point is to consider first order deviations in the network activity from the quiescent state.  Here $\mathbf{C}(\tau)\approx\bm{\Delta}(\tau)$ where $\bm{\Delta}(\tau)$ is the autocorrelation vector of the activities with elements $\Delta_d (\tau)=\left\langle\xi_d(t)\xi_d(t+\tau)\right\rangle$. By invoking Eq.~(\ref{eq:mfield}) we have
\begin{equation}
\mathbf{H}(\tau) = \left(1 - \frac{d^2}{d\tau^2}\right)\bm{\Delta}(\tau).
\label{eq:HD}
\end{equation}
Substituting Eq.~(\ref{eq:HD}) into Eq.~(\ref{eq:matrix_form}) leads to an equation of motion of a particle in a harmonic potential for $\bm{\Delta}(\tau)$:
\begin{equation}
\frac{d^2\bm{\Delta}(\tau)}{d\tau^2} = \left(\mathbb{I} - \mathbf{M}\right)\bm{\Delta}(\tau).
\end{equation}

\begin{figure}[b]
\begin{centering}
\includegraphics[width = 0.98\columnwidth]{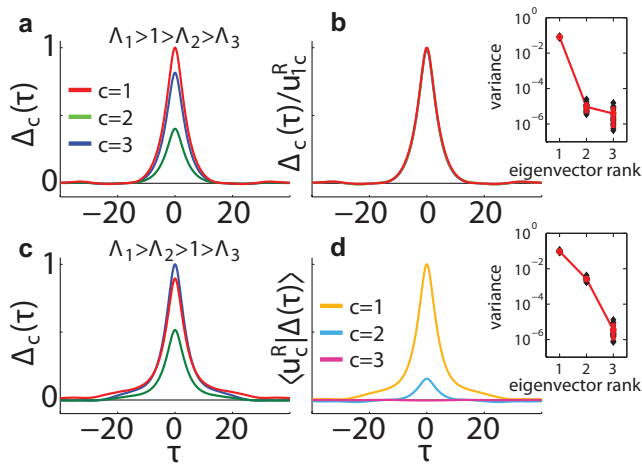}
    \caption{Autocorrelation modes. Example networks have $N = 1200$ and $3$ equally sized groups with $\alpha, \mathbf{g}$ such that $\mathbf{M}$ is symmetric. (a) When $D^{\star} = 1$, autocorrelations maintain a constant ratio independent of $\tau$. (b) Rescaling by the components $u^R_{1c}$ merges the autocorrelation functions. (c) When $D^{\star} = 2$, the autocorrelation functions are linear combinations of two autocorrelation ``modes'' that decay on different timescales.  Projections of these functions $\langle u^R_c|\bm {\Delta}(\tau)\rangle$ are shown in (d). Only projections on $|u^R_1\rangle, | u^R_2\rangle$ are significantly different from $0$.  Insets show the variance of $\bm{\Delta}\left(\tau\right)$ projected on $\left|u^R_c\right\rangle$ averaged over $20$ networks in each setting.}
    \label{fig:ac_modes}
\end{centering}
\end{figure}

The shape of the multivariate potential depends on the eigenvalues of $\mathbf{M}$. The first bifurcation (assuming the elements of $\mathbf{M}$ are scaled together) occurs when $\Lambda_1=1$, in the direction parallel to the leading eigenvector. Physical solutions should have $\|\bm{\Delta}(\tau)\| < \infty$ as $\tau\rightarrow\infty$ because $\bm{\Delta}(\tau)$ is an autocorrelation function. When all eigenvalues of $\mathbf{M}$ are smaller than $1$ the trivial solution $\bm{\Delta}(\tau) = 0$ is the only solution (in the neighborhood of $x_i(t)=0$ where our approximation is accurate). At the critical point ($\Lambda_1 = 1$) a non trivial solution appears, and above it finite autocorrelations lead to chaotic dynamics in the full system \footnote{Establishing the existence of a positive Lyapunov exponent requires analysis of the full self consistency equations and cannot be done using the linearized approximation}.

Recall the dependence of the connectivity parameters and the critical point at which the network transitions to chaos. In the single and multiple cell-type networks this transition occurs when a finite mass of the spectral density of $\mathbf{J}$ has real part $>1$. Thus, when $\Lambda_1 =1$ the eigenvalues of $\mathbf{J}$ are bounded in the unit circle $r=1$. For networks with cell-type independent connectivity, $\Lambda_1=g^2$ and $r=g$. Requiring continuity, this implies that the circle that bounds the eigenvalue density of $\mathbf{J}$ has radius
\begin{equation}
    r(\alpha,\mathbf{g}) = \sqrt{\Lambda_1} = \sqrt{\max\left[\lambda(\mathbf{M})\right]}.
\label{eq:radius}
\end{equation}
We have verified Eq.~(\ref{eq:radius}) using numerical simulations (Fig. \ref{fig:example}) for a number of different matrix configurations. Strikingly, $r$ is qualitatively different from the mean synaptic gain $\bar{g}$ (Fig. \ref{fig:example}a,b). The inequality $\sqrt{\Lambda_1}\neq\bar{g}$ is a signature of the block structured variances. It is not observed in the case where the variances have columnar structure \cite{Rajan2006}, when $\text{rank}\{\mathbf{M}\}=1$ \cite{Wei2012,Ahmadian2013}, or when the $J_{ij}$'s are randomly permuted.

\paragraph*{Autocorrelation modes.}
Next we analyze the network dynamics above the critical point. In the chaotic regime the persistent population-level activity is determined by the structure of the matrix $\mathbf{M}$. Consider the decomposition $\mathbf{M}=\sum_{c=1}^D\Lambda_c|u^R_c\rangle\langle u^L_c|$ where $|u^R_c\rangle, \langle u^L_c|$ are the right and left eigenvectors ordered by the real part of their corresponding eigenvalues $\Re\{\Lambda_c\}$, satisfying $\langle u^L_c|u^R_d\rangle = \delta_{cd}$. By analogy to the analysis of the scalar self consistency equation in \cite{Sompolinsky1988} we know that the trivial solution to Eq.~(\ref{eq:matrix_form}) is unstable in the subspace $\mathcal{U}_{\mathbf{M}} = \text{span}\{|u^R_1\rangle, \dots,|u^R_{D^{\star}}\rangle\}$, where $D^{\star}$ is the number of eigenvalues of $\mathbf{M}$ with real part $>1$. In that subspace the solution to
Eq.~(\ref{eq:matrix_form}) is a linear combination of $D^{\star}$ different autocorrelation functions.  Conversely, in the $D-D^{\star}$ dimensional orthogonal complement subspace $\mathcal{U}_{\mathbf{M}}^{\perp}$ the trivial solution is stable. As a consequence, the vectors $\mathbf{H}(\tau),\bm{\Delta}(\tau)$ are restricted to $\mathcal{U}_{\mathbf{M}}$ and their projection on any vector in $\mathcal{U}_{\mathbf{M}}^{\perp}$ is $0$.

In the special case $D^{\star} = 1$ only one eigenvalue of $\mathbf{M}$ has a real part $>1$, and the activity of neurons in all groups follows the same autocorrelation function. The scaling is determined by the components $u^R_{1c}$ of the leading right eigenvector of $\mathbf{M}$ (see Fig.~\ref{fig:ac_modes}a,b): $\Delta_c(\tau)/\Delta_d(\tau) = u^R_{1c}/u^R_{1d}$. In general $D^\star$ can be larger than $1$. In Fig~\ref{fig:ac_modes}c,d we show an example network with three cell-types and $D^{\star}=2$.
Note that for asymmetric $\mathbf{M}$, $|u^R_c\rangle$ are not orthogonal and $\mathcal{U}_{\mathbf{M}}^\perp$ is spanned by the left rather than
the right eigenvectors: $\mathcal{U}_{\mathbf{M}}^{\perp}=\text{span}\{\langle u^L_{D^{\star} + 1}|, \dots, \langle u^L_{D}|\}$.


\paragraph*{Universality and sparsity.}
\begin{figure}[t]
\begin{centering}
\includegraphics[width=1\columnwidth]{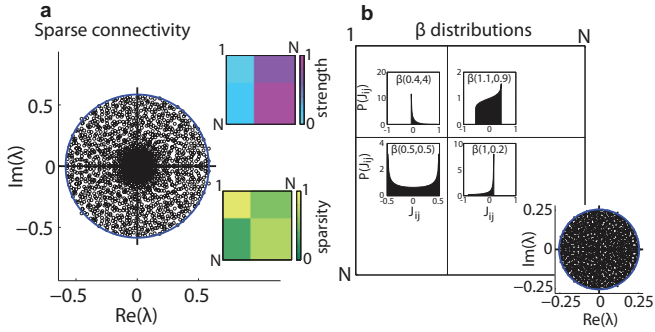}
    \caption{Universality and sparse connectivity. (a) Our results extend to sparse connectivity matrices, an example matrix with non-Gaussian element distributions. The formula for the radius (blue circle) is in agreement with the numerical results.  Insets shows average synaptic strengths, $\mathbf{G}$ (top) and the sparsity levels (bottom). (b) In each block the elements of $\mathbf{J}$ were drawn from a centered $\beta$ distribution with different parameters leading to skewed and bimodal distributions.}
    \label{fig:universality}
\end{centering}
\end{figure}

Until now we have discussed connectivity matrices with elements drawn from Gaussian distributions. However Girko's
circular law \cite{Girko1984,Bai1997} is universal, meaning that the spectral density of connectivity matrices describing single cell-type networks depends only on the second moment of the matrix entry distribution \cite{Tao2010} (as long as the mean remains zero). This suggests that our results for matrices with block structured variances, extend to non-Gaussian distributions, provided that $\langle J_{ij}\rangle = 0$ and $N \langle J_{ij}^2\rangle = g^2_{c_id_j} < \infty$. Using numerical simulations, we have verified that Eq.~(\ref{eq:radius}) holds for a number of non-Gaussian matrix element distributions, including networks where connection strengths were taken from sparse and $\beta$ distributions (Fig.~\ref{fig:universality}) \footnote{In the sparse example, $s_{cd}$ is the fraction of nonzero elements, randomly drawn from a Gaussian distribution with variance $g^2_{cd}/N$. The block-wise variance  is therefore $s_{cd}g^2_{cd}/N$, and eigenvalues are bounded by a circle with radius calculated using $M_
{cd}=\alpha_d s_{cd}g^2_{cd}$.}.

\paragraph*{Applications.}
\begin{figure}
\begin{centering}
\includegraphics[width = 1\columnwidth]{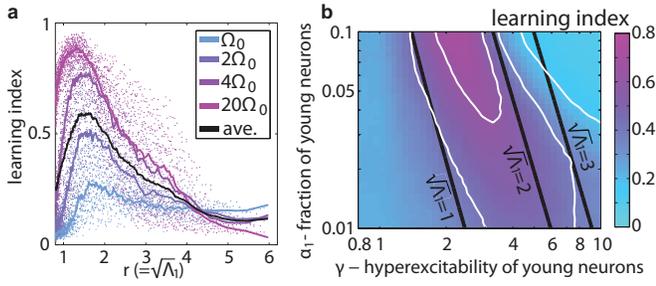}
    \caption{Application to neurogenesis. (a) The learning index $l_{\Omega}\left (\alpha_1,\gamma\right)$ for four pure frequency target functions ($\Omega_0 = \pi/120$) plotted as a function of the radius $r = \sqrt{\Lambda_1}$. The training epoch lasted approximately $100$ periods of the target signal. The radius is a good coordinate to describe the learning capacity. (b) The same data, averaged over the target frequencies, plotted in the $\gamma-\alpha_1$ plane. Contour lines of $l_{\Omega}\left(\alpha_1, \gamma \right)$ (white) and of $\sqrt{\Lambda_1}$ (black) coincide approximately in the region where $l_{\Omega}$ peaks, supporting our conclusion that learning is modulated in parameter space primarily by $\sqrt{\Lambda_1}$, the effective gain of the network.}
    \label{fig:neurogenesis}
\end{centering}
\end{figure}
We now illustrate how these theoretical results can give insight into a perplexing question in computational neuroscience - how can a small number of neurons have a large effect on the representational capacity of the whole network? In adults, newborn neurons continuously migrate into the existing neural circuit in the hippocampus and olfactory bulb regions \cite{Zhao2008}. Impaired neurogenesis results in strong deficits in learning and memory. This is surprising since the young neurons, although hyperexcitable, constitute only a very small fraction ($<0.1$) of the total network. To better understand the computational role young neurons may play, we analyzed dynamics of a network with $D=2$ groups of neurons.  The first group consists of the young neurons, so its size is significantly smaller than the second group which consists of the mature neurons ($\alpha_1 \ll \alpha_2$). In our model, the connectivity within the existing neural circuit (group 2) is such that by itself that subnetwork would be in the 
quiescent state: $g_{22}=1-
\epsilon<1$. To model the increased excitability of the young neurons all connections of these neurons were set to a larger value: $g_{12}=g_{21}=g_{11}=\gamma>1-\epsilon$.

We analyzed the network's capacity to accurately reproduce a target output pattern. The activity of the neurons serves as a ``reservoir'' of waveforms from which the target signal $f(t)$ is composed. We used the learning algorithm in \cite{Sussillo2009} to find the vector $\mathbf{w}$ such that $z(t)= \sum_{i=1}^N w_i\phi_i(t)=f(t)$, where the modified dynamics have $J_{ij}\rightarrow J_{ij}+u_iw_j$ and $\mathbf{u}$ is a random vector with entries of $O(1)$. The single group network does this well when its synaptic gain is $g\approx 1.5$ \cite{Sussillo2009}, such that its initial activity is in the weakly chaotic regime. For simplicity we choose purely periodic target functions $f(t) = A\sin(\Omega t)$. We define the learning index as $l_{\Omega} = |\tilde{z}(\Omega)|/\int |\tilde{z}(\omega)|^2d\omega$, with $\tilde{z}(\omega)$ being the Fourier transform of $z(t)$ generated by the network after the learning epoch.

For fixed $\epsilon = 0.2$ and $N = 500$ we computed $l_{\Omega}(\alpha_1,\gamma)$ and found that for this family of networks $\sqrt{\Lambda_1}$ plays a role equivalent to that of $g$ in the single group network. Performance is optimal for $\sqrt{\Lambda_1}\approx 1.5$, and networks with different structure perform similarly as long as they have similar values of $\Lambda_1$ (Fig.~\ref{fig:neurogenesis}). These results demonstrate that a small group of neurons could place the overall network in a state conducive to learning. Importantly, because of the block structured connectivity, the effective gain is larger than the average gain ($\sqrt{\Lambda_1}>\bar{g}$), suggesting that modulating the synaptic gain can carry a larger effect on the learning capacity of the multiple cell-type network compared to what one may expect based on changes in mean connectivity gain.


It is worth noting that typically outgoing connections from any given neuron are all positive or all negative, obeying Dale's law \cite{Eccles1976}. Within random networks, this issue was addressed by Rajan and Abbott \cite{Rajan2006} who studied a model where columns of $\mathbf{J}$ are separated to two groups, each with its offset and element variance. They computed the bulk spectrum by imposing a ``detailed balance'' constraint, where the sum of incoming connections to each neuron is exactly $0$ \cite{Rajan2006,Wei2012}. The distribution of outliers which appear when this constraint is lifted was computed by Tao \cite{Tao2013}. The dynamics of networks with cell-type-dependent connectivity that is offset to respect Dale's law were addressed in \cite{Cabana2013} with some limitations, and remain an important problem for future research.

Ultimately, neural network dynamics need to be considered in relation to external inputs. The response properties of networks with one cell-type have been recently worked out \cite{Rajan2010,Toyoizumi2011}.  The analogy between the mean field equations for the single and multi-group cases suggests that our results can be used to understand the non-autonomous behavior of multiple cell-type networks.

The authors would like to thank Larry Abbott for his support, including comments on the manuscript, and Ken Miller for many useful discussions.  This work was supported by NIH grant (R01EY019493) and NSF Career award (IIS 1254123). MS was supported by the Gatsby Foundation.

%

\end{document}